\documentclass[12pt]{iopart}
\usepackage{times}
\usepackage{graphicx}
\usepackage[pdftex]{color}

\expandafter\let\csname equation*\endcsname\relax
\expandafter\let\csname endequation*\endcsname\relax

\usepackage{amsmath}
\usepackage{bbm}
\usepackage{bm}
\DeclareBoldMathCommand{\bfmu}{\mu}
\DeclareBoldMathCommand{\bfnabla}{\nabla}

\newcommand{\currentJ}{\mathbf{j}}

\renewcommand{\vec}[1]{\mathbf{#1}}

\begin{document}

\title{%
Fully covariant radiation force on a polarizable particle%
}
\author{%
Gregor Pieplow and
Carsten Henkel%
\footnote[3]{henkel@uni-potsdam.de}}
\address{Institute of Physics and Astronomy, Universit\"at Potsdam,
Karl-Liebknecht-Str. 24/25, 14476 Potsdam,
Germany}%

\begin{abstract}
\noindent
The electromagnetic force on a polarizable particle is calculated in a covariant
framework. Local equilibrium temperatures for the electromagnetic field
and the particle's dipole moment are assumed, using a relativistic formulation
of the fluctuation-dissipation theorem. Two examples illustrate
radiative friction forces: a particle moving through a homogeneous radiation
background and above a planar interface. Previous results for arbitrary
relative velocities are recovered in a compact way. 
\end{abstract}

\date{28 Sep 2012}

\section{Introduction}

Friction is ubiquitous in everyday life and omnipresent in almost every 
mechanical system, but still very difficult to grasp at the level of 
elementary forces. For example, internal friction in fluids (viscosity) or 
contact friction between two solids can be attributed to electromagnetic 
interactions who couple individual atoms. Building a bridge from the microscopic
realm to the macroscopic world is a challenge because 
of the huge diversity at the atomistic level: surface reconstruction, adsorbates,
roughness \ldots~.
A common feature of friction forces, however, is the conversion of directed 
motion into thermal motion or material excitations. This feature can be studied
with the help of simple models at the fundamental level. In recent years,
studies of moving objects have been developed from first principles: consider,
for example, a flat surface separated by vacuum from another body in
constant parallel motion. On a length scale of a few nanometers, the 
electromagnetic interactions between the two can be characterized by a
few macroscopic parameters (refractive index, conductivity, surface 
impedance \ldots). In this framework, consistent quantum field theories
have been formulated \cite{Barnett1997, Eberlein2006a, Buhmann2007a} 
that in principle can take full advantage of 
Lorentz invariance. An example is the reflection of light from a moving plate
that is evaluated by transforming the incident field into the plate's rest (or
co-moving) frame and back. In the same spirit, this co-moving frame is
a natural candidate for local thermodynamic equilibrium, at least for
macroscopic objects. (For a discussion on the transformation law of
temperature and its dependence on the definition of relative motion,
see Ref.\cite{Biro2010}.)
From the viewpoint of relativistic thermodynamics \cite{Neugebauer_book, Melrose_v1},
the two situations of bodies in relative motion or fixed at different 
temperatures indeed represent very similar non-equilibrium settings.

In this paper, we construct a fully covariant formulation of radiation-induced
forces on a small neutral particle. We consider stationary, non-equilibrium
motion at arbitrary speed parallel to a planar surface. 
Local temperatures are assigned to the particle and the
surface, in their respective rest frames. Covariance is maintained from the
beginning and expresses in a compact form the transformation properties
of the electromagnetic field, the material polarization, and the particle's
dipole moments. The fluctuation-dissipation theorem that determines the 
spectra of thermal fluctuations is formulated in local (co-moving) frames 
of particle or surface, respectively, which provides a natural link to 
relativistic thermodynamics. We check our general expression for the
radiation force by specializing to motion through the 
blackbody radiation field and to a particle above a dielectric surface. In the
two cases, we use different gauges, but come to results fully consistent with
previous work \cite{Mkrtchian2003, dedkov2003relativistic}. We believe that the present formulation is useful
because it is compact and flexible, and illustrates the assumptions 
behind the macroscopic
quantum field theory in a physically transparent way. This may pave the way
to interpret electromagnetic friction phenomena that have attracted some
interest over the last years \cite{Barton1996a, pendry1997shearing, volokitin2008theory, philbin2009no, Scheel2009moving}.

The outline is as follows: the covariant framework is constructed
in Sec.~\ref{s:framework}, resulting in the fluctuation-dissipation theorem 
for the electromagnetic field and the particle's dipole moment in
Sec.~\ref{s:local-FDT}. The force is split in two contributions that
can be attributed to radiation reaction and vacuum fluctuations \cite{Cohen1982, Meschede1990},
both are given in general form in Sec.~\ref{s:general-formula}. 
We specialize to blackbody friction in Sec.~\ref{s:bb-friction} and to
radiation forces above a surface in Sec.~\ref{s:surface-friction}.

\section{Covariant framework}
\label{s:framework}

\subsection{Polarization and force density}

In this part we introduce a covariant expression for the force acting on a
polarizable body. We start with some basic identities from electrodynamics: 
polarization and magnetization fields $\vec{P}$ and $\vec{M}$ are defined from
\begin{equation}
	\rho = - \vec{\nabla} \cdot \vec{P}
	, 
	\qquad
	\currentJ = \partial_t \vec{P} + \vec{\nabla} \times \vec{M}
	\label{eq:def-P-and-M}
\end{equation}
where $\rho$ is the charge density and $\currentJ$ the spatial current density.
In terms of the four vector $(j^\mu) = ( \rho, \currentJ ) = 
(\rho, j^1, j^2, j^3)$:
\begin{equation}
	j^\mu = \partial_{\nu} M^{\nu\mu}
	\label{eq:4-current}
\end{equation}
where
$M^{\nu\mu}$ is the polarization tensor with the matrix representation
\begin{equation}
	(M^{\nu\mu}) = \left(
	\begin{array}{cccc}
	0 & P^1 & P^2 & P^3
	\\
	- P^1 & 0 & M^3 & - M^2
	\\
	- P^2 & -M^3 & 0 & M^1
	\\
	- P^3 & M^2 & -M^1 & 0
	\end{array} 
	\right)
	~.
	\label{eq:M-matrix}
\end{equation}
Its antisymmetry ensures charge conservation. The electromagnetic
force density
\begin{equation}
	\vec{f} = \rho \vec{E} + \currentJ \times \vec{B}
	,
	\qquad
	\label{eq:4-force-and-current}
\end{equation}
is part of the 4-vector
\begin{equation}
	f_\mu = F_{\mu\nu} j^\nu
\end{equation}
where the field strength (Faraday) tensor has components
\begin{equation}
	(F_{\mu\nu}) = \left(
	\begin{array}{cccc}
	0 & E^1 & E^2 & E^3
	\\
	- E^1 & 0 & B^3 & - B^2
	\\
	- E^2 & - B^3 & 0 & B^1
	\\
	- E^3 & B^2 & -B^1 & 0
	\end{array} 
	\right)
	\label{eq:F-matrix}
\end{equation}
Pulling these relations together, we can write the covariant force density 
in terms of the field and polarization tensors
\begin{equation}
	f_\mu = F_{\mu\nu} \partial_\kappa M^{\kappa\nu}
	\label{eq:4-force-and-M}
\end{equation}
The total force on a body is found by integrating %~(\ref{eq:4-force-and-M}) 
over a volume containing the body. Up to surface terms in this integral,
Eq.(\ref{eq:4-force-and-M}) is equivalent to the Einstein-Laub formula 
for the force on polarizable matter \cite{Einstein1908b,Jackson,Mansuripur2012}.

In the following, we focus on the situation that the force $f_\mu$
[Eq.(\ref{eq:4-force-and-M})] arises from fluctuations of the field
and of the material polarization. In the spirit of perturbation theory,
we split the polarization, for example, into
\begin{eqnarray}
M^{\mu\nu}(x) & = &
	M^{\mu\nu}_{\rm fl}(x) + M^{\mu\nu}_{\rm in}(x)~,
\label{eq:split-M}
\end{eqnarray}
where the first term ``fl'' describes the free fluctuations of the electric
and magnetic dipole moments, while the second term ``in'' (= induced)
gives the response
to an exterior field. A similar split for the fields yields an average force
density
\begin{equation}
f_{\mu}( x ) = 
\langle F^{\rm fl}_{\mu\nu}(x)\partial_{\sigma}M_{\rm in}^{\sigma\nu}(x) 
\rangle_F
+ 
\langle
F^{\rm in}_{\mu\nu}(x)\partial_{\sigma}M_{\rm fl}^{\sigma\nu}(x)
\rangle_A
~,
\label{eq:Dforcesplitting}
\end{equation}
where the subscripts $F$ and $A$ denote the canonical average with respect 
to the field and atom Hamiltonians. In the first order of perturbation theory,
these averages are evaluated with local equilibrium temperatures
$T_F$ and $T_A$. To make the two terms real-valued, the operator products
must be symmetrized [see after Eq.(\ref{eq:FD-dipole-nrel})], as discussed
in Ref.\cite{Cohen1982}.
In the following sections, we spell out the linear response functions and
the fluctuation spectra, respectively.

\subsection{Response functions}

\subsubsection{Polarizability.}

For simplicity, we focus on a pointlike particle (``atom'') at position $\vec{x}_A$
with an electric dipole polarizability which is often the dominant response.
The particle carries an electric dipole
moment that responds to the electric field vector,
\begin{equation}
	\vec{d}( t ) = \alpha
	\vec{E}( t, \vec{x}_A )
	\label{eq:def-alpha}
\end{equation}
where $\alpha$ is called the polarizability whose frequency dependence
(dispersion) is taken into account in Eq.(\ref{eq:alpha-of-omega}) below.
We assume an isotropic response. 
Introducing the 4-velocity $u^\mu$ tangent to the particle's worldline
$x_A$, we define the covariant magnetization density 
\begin{equation}
	M^{\mu\nu}( x ) = \left[ u^\mu( t ) d^\nu( t ) - 
	d^\mu( t ) u^\nu( t ) \right] \delta( \vec{x} - \vec{x}_A( t ) )
	\label{eq:def-M-for-dipole}
\end{equation}
The component of $d^\mu$ parallel to $u^\mu$ can 
be chosen arbitrarily, it drops out from this construction. 
The atom responds to the
electric field in the co-moving frame so that the covariant form of
Eq.(\ref{eq:def-alpha}) becomes 
\begin{equation}
	d^\mu( x_A ) = \alpha g^{\mu\kappa}F_{\kappa\lambda}( x_A ) 
		u^\lambda
	\label{eq:4-dipole}
\end{equation}
where $g^{\mu\kappa}$ is the metric tensor. This 4-vector
$d^\mu$ is indeed perpendicular to $u^\mu$. 
A similar construction can be given for
the magnetic polarizability. Making the approximation
that the atom's worldline is inertial (constant $u^\mu$), we get
in Fourier space ($k \cdot x = k_\mu x^\mu$)
\begin{eqnarray}
	\hspace*{-10mm}
	M^{\mu\nu}( x ) &=& 
	\int\!\frac{ {\rm d}^4 k }{ (2\pi)^4 }
	{\rm e}^{ - {\rm i} k \cdot x }
	M^{\mu\nu}( k ) =
	\int\!\frac{ {\rm d}^4 k }{ (2\pi)^4 }
	\frac{ {\rm d}^4 h }{ (2\pi)^4 }
	{\rm e}^{ - {\rm i} k \cdot x }
	\alpha^{\mu\nu\kappa\lambda}( k, h )
	F_{\kappa\lambda}( h )
\label{eq:alpha-of-omega}
\\
	\hspace*{-10mm}
	\alpha^{\mu\nu\kappa\lambda}( k, h ) &=&
	2\pi \delta( u \cdot ( k - h ) )
	\alpha( u \cdot h )
	\left\{ u^{\mu} g^{\nu\kappa} u^{\lambda} + 
	u^{\nu} g^{\mu\lambda} u^{\kappa}
	\right\}
	\, {\rm e}^{ {\rm i} ( k  - h ) \cdot x_A }
\label{eq:4-rank-alpha}
\end{eqnarray}
Here we have restored dispersion via the argument of the polarizability: the
quantity $\omega'_A = u \cdot h$ plays the role of the frequency of a wavevector
component $h_\mu$ of the applied field, as seen in the atom's rest frame. This 
describes in particular the first- and second-order Doppler shifts. The dipole
radiates at the same frequency as the applied field (linear response), hence the 
$\delta( u \cdot ( k - h ) )$. We work
in this paper with retarded response functions: the 
dipole responds to the electric field in its past and therefore, 
$\alpha( \omega'_A )$ is analytic in the upper half-plane of complexified 
frequencies $\omega'_A$. For the radiation force~(\ref{eq:Dforcesplitting}),
Eq.(\ref{eq:alpha-of-omega}) gives the polarization
$M^{\mu\nu}_{\rm in}$ induced by the fluctuating field
$F_{\kappa\lambda}^{\rm fl}$.

\subsubsection{Green function.}

It is well known that for the electromagnetic field, the vector potential
created by a source current can be found with the help of a Green function,
\begin{equation}
	A_\mu( x ) = \int\!{\rm d}^4y \, {\cal G}_{\mu\nu}( x, x' ) j^{\nu}( x' )
	\label{eq:def-Green-tensor}
\end{equation}
In free space, for example, we have, adopting the Feynman gauge
($k^2 = k \cdot k = \omega^2 - \vec{k}^2$, we set $c = \varepsilon_0 = 1$)
\begin{equation}
	\mathcal{G}_{\mu\nu}( x, x' ) = 
	\int\!\frac{ {\rm d}^4 k }{ (2\pi)^4 }
	{\rm e}^{ - {\rm i} k \cdot ( x - x' ) }
	\frac{ - g_{\mu\nu} }{ 
		k^2 + {\rm i} 0 \mathop{\mathrm{sgn}}{\omega} }
\label{eq:Feynman-Green-free-space}
\end{equation}
In the general case, translation invariance in time or space can only hold
under special circumstances, so we adopt the Fourier expansion
\begin{equation}
	\mathcal{G}_{\mu\nu}( x, x' ) =
	\int\!\frac{ {\rm d}^4k }{ (2\pi)^4 }
	\int\!\frac{ {\rm d}^4h }{ (2\pi)^4 }
	{\rm e}^{ - {\rm i} (k \cdot x - h \cdot x') }
	\mathcal{G}_{\mu\nu}( k, - h )
	~.
\label{eq:Fourier-Gmn}
\end{equation}
If we represent the current density in terms of the polarization 
$M^{\kappa\lambda}$ [Eq.(\ref{eq:4-current})], we get the field amplitude from
\begin{eqnarray}
	F_{\mu\nu}( x ) &=& 
	\int\!\frac{ {\rm d}^4k }{ (2\pi)^4 }
	\int\!\frac{ {\rm d}^4h }{ (2\pi)^4 }
	{\rm e}^{ - {\rm i} k \cdot x }
	{\cal G}_{\mu\nu\kappa\lambda}( k, - h )
	M^{\kappa\lambda}( h )	
\label{eq:def-Green-4-tensor}
\\
	{\cal G}_{\mu\nu\kappa\lambda}( k, - h ) &=&
	k_{\mu} \mathcal{G}_{\nu\kappa}( k, - h ) h_{\lambda} + 
	k_{\nu} \mathcal{G}_{\mu\lambda}( k, - h ) h_{\kappa}
\label{eq:link-Green-tensors}
\end{eqnarray}
where the antisymmetry of $M^{\kappa\lambda}$ was used. 
Note the formal
analogy to the fourth rank polarization tensor
$\alpha^{\mu\nu\kappa\lambda}$ [Eq.(\ref{eq:4-rank-alpha})].
For the radiation force~(\ref{eq:Dforcesplitting}), we shall use 
Eq.(\ref{eq:def-Green-4-tensor}) to express the field
$F_{\mu\nu}^{\rm in}$ radiated by the fluctuating polarization
$M^{\kappa\lambda}_{\rm fl}$.

\subsection{Fluctuation spectra}
\label{s:local-FDT}

To evaluate the radiation force~(\ref{eq:Dforcesplitting}), we need
correlation functions of the fluctuating polarization and fields. These are
provided by the fluctuation-dissipation theorem, assuming thermal states
for atom and field.

\subsubsection{Dipole and polarization.}

The non-relativistic form of the dipole correlation function 
is~\cite{Landau_v9,Rytov_v3} ($i, j$ are spatial components)
\begin{equation}
	\langle d^i( \omega ) , d^j( \omega' ) \rangle_A = 
	2\pi \hbar \delta( \omega + \omega' ) 
	\delta^{ij}
	\coth\big( \frac{ \hbar \omega }{ 2 k_{\rm B} T_A } \big)
	\mathop{\mathrm{Im}}\alpha( \omega ) 
	\label{eq:FD-dipole-nrel}
\end{equation}
where the operator product 
is symmetrized: 
$\langle B , C \rangle_A = \frac12 \langle B C + C B \rangle_A =
\frac12 {\rm tr}\left\{ \rho_A ( B C + C B ) 
\right\}$ with the equilibrium density operator $\rho_A$.
$\mathop{\mathrm{Im}}\alpha( \omega )$ describes the spectral distribution 
of the atomic oscillator strength that may also depend on the atomic 
temperature $T_A$. In the relativistic formulation, the 
inverse temperature  becomes a time-like
4-vector $\beta_A^\mu = (\hbar / k_{\rm B} T_A) u^\mu$ tangent 
to the atom's worldline~\cite{Neugebauer_book, Melrose_v1}. 
We recover Eq.(\ref{eq:FD-dipole-nrel}) in the spacelike hypersurface
perpendicular to $u^\mu$ when the following correlation function 
for the polarization field $M^{\mu\nu}( x )$ localized on the atom
is assumed (written in 4D Fourier space)
\begin{eqnarray}
	\hspace*{-15mm}
	\langle M^{\mu\nu}_{\rm fl}( k ) ,
	M^{\kappa\lambda}_{\rm fl}( h ) \rangle &=&
	2\pi \hbar \delta( u \cdot ( k + h ) ) 
	\Gamma^{[\mu\nu][\kappa\lambda]}
	\coth\big( \frac{ \beta_A \cdot k }{ 2 } \big)
	\mathop{\mathrm{Im}}\alpha( u \cdot k ) 
	\,{\rm e}^{ {\rm i} (k + h) \cdot x_A }
\label{eq:FD-dipole-cov}
\\
	\Gamma^{[\mu\nu][\kappa\lambda]} &=&
	u^{[\mu} g^{\nu] [\kappa} u^{\lambda]}	
\end{eqnarray}
Here, the square brackets are denoting odd combinations of paired indices
\begin{equation}
	u^{[\mu} g^{\nu] [\kappa} u^{\lambda]}	
=
	u^{\mu} g^{\nu \kappa} u^{\lambda}	
-
	u^{\nu} g^{\mu \kappa} u^{\lambda}	
-
	u^{\mu} g^{\nu \lambda} u^{\kappa}	
+
	u^{\nu} g^{\mu \lambda} u^{\kappa}	
	\label{eq:anti-symmetrize}
\end{equation}
and ensure the antisymmetry of the fluctuating polarization tensor.
For a non-inverted atom, the absorption spectrum 
$\sim \omega \, \mathop{\mathrm{Im}}\alpha( \omega )$ is positive for
all frequencies; this property is inherited by the correlation 
spectrum~(\ref{eq:FD-dipole-cov}).

\subsubsection{Field correlations.}

The correlations of the electromagnetic fields $\vec{E}$ and $\vec{B}$
are well known at thermal equilibrium in the rest
frame \cite{Rytov_v3, Agarwal1975}
\begin{eqnarray}
	\hspace*{-20mm}
	\langle {E}_i(\omega,\vec{x}) , {E}_j(\omega',\vec{x}')\rangle_{F}
	&=&
	\begin{aligned}[t]
	{\rm i}\hbar\pi\delta(\omega+\omega') 
	& \coth \frac{ \beta_F \omega }{ 2 } \times
\\ 
	& \omega^2 
\left[\mathcal{G}_{ij}(\omega;\vec{x},\vec{x}') - {\mathcal{G}_{ji}}^*(\omega;\vec{x}',\vec{x})\right]~,
	\end{aligned}
\label{eq:DfdtEijort}
\\
	\hspace*{-20mm}
	\langle {E}_i(\omega,\vec{x}) , {B}_j(\omega',\vec{x}')\rangle_{F}
	&=&
	\begin{aligned}[t]
	-\hbar\pi \delta(\omega+\omega') 
	& \coth \frac{ \beta_F \omega }{ 2 } \times
\\ 
	& \omega\, \epsilon_{klj} \partial'_{k} 
\left[\mathcal{G}_{il}(\omega;\vec{x},\vec{x}') - {\mathcal{G}_{li}}^*(\omega;\vec{x}',\vec{x})\right]~,
	\end{aligned}
\label{eq:DfdtSijort}
\\
	\hspace*{-20mm}
	\langle {B}_i(\omega,\vec{x}) , {B}_j(\omega',\vec{x}')\rangle_{F}
	&=&
	\begin{aligned}[t]
	{\rm i}\hbar\pi \delta(\omega+\omega')  
	& \coth \frac{ \beta_F \omega }{ 2 } \times
\\ 
	& \epsilon_{kli}\epsilon_{mnj}\,\partial_{k} \partial'_{m}
	\left[\mathcal{G}_{ln}(\omega;\vec{x},\vec{x}') - {\mathcal{G}_{nl}}^*(\omega;\vec{x}',\vec{x})\right]~.
	\end{aligned}
\label{eq:Dfdthijort}
\end{eqnarray}
where now $\beta_F$ is the inverse field temperature. In the presence of
macroscopic bodies, we assume that the field relaxes to their temperature.
The Green tensor in Eqs.(\ref{eq:DfdtEijort}, \ref{eq:DfdtSijort}, 
\ref{eq:Dfdthijort}) is calculated from the Kubo formula
\begin{equation}
	\mathcal{G}_{ik}(\omega;\vec{x},\vec{x'}) = - \frac{ {\rm i} }{ \hbar }
	\int_{t'}^{\infty}dt\,{\rm e}^{i\omega (t - t')}
	\langle{A}_i(t,\vec{x}){A}_k(t',\vec{x}') - 
	{A}_k(t',\vec{x'}){A}_i(t,\vec{x})\rangle_{F}
	~.
\label{eq:Kubo-Green-tensor}
\end{equation}
in the Dzyaloshinskii gauge (zero scalar potential). Note that this
definition assumes that the field is stationary in the chosen frame, i.e.\
correlations $\langle{A}_i( x ){A}_k( x' )\rangle_{F}$ of the
vector potential depend only on the time difference $t - t'$. 
The integration is over retarded times $t > t'$ in this frame. It is a
fundamental result of linear response theory that the correlation function
of Eq.(\ref{eq:Kubo-Green-tensor}) also provides the Green function for
the vector potential, i.e., the kernel in Eq.(\ref{eq:def-Green-tensor}).

We proceed to combine the field spectra into a covariant formulation.
In a first step, we complete Eqs.(\ref{eq:DfdtEijort}--\ref{eq:Dfdthijort}) by
performing a spatial Fourier transformation, yielding
($h = ( \omega', \vec{h} )$)
\begin{eqnarray}
	\hspace*{-20mm}
	&\begin{aligned}
\langle {E}_i( k ) , {E}_j( h )\rangle_{F}
= {\rm i} \hbar\pi\delta(\omega+\omega') 
& \coth \frac{\beta_F \omega }{ 2 } \times
\\ 
& \omega^2 \left[\mathcal{G}_{ij}(\omega,\vec{k},\vec{h})-\mathcal{G}_{ji}^*(\omega,-\vec{h},-\vec{k})\right]~,
\end{aligned}
\label{eq:DfdtEij}
\\
	\hspace*{-20mm}
&\begin{aligned}
\langle {E}_i( k ) , {B}_j( h )\rangle_{F}
= - {\rm i} \hbar\pi \delta(\omega+\omega') 
& \coth \frac{\beta_F \omega }{ 2 } \times
\\ 
& \epsilon_{jkl} \omega h_k \left[\mathcal{G}_{il}(\omega,\vec{k},\vec{h})-\mathcal{G}_{li}^*(\omega,-\vec{h},-\vec{k})\right]~,
\end{aligned}
\label{eq:DfdtSij}
\\
	\hspace*{-20mm}
&\begin{aligned}
\langle {B}_i( k ) , {B}_j( h )\rangle_{F}
= - {\rm i} \hbar\pi  \delta(\omega+\omega')
& \coth \frac{\beta_F \omega }{ 2 } \times
\\ 
& \epsilon_{ikl}\epsilon_{jmn}\,k_k h_m \left[\mathcal{G}_{ln}(\omega,\vec{k},\vec{h})-\mathcal{G}_{nl}^*(\omega,-\vec{h},-\vec{k})\right]
~,
\end{aligned}
\label{eq:Dfdthij}
\end{eqnarray}
using for the Green tensor the expansions
\begin{eqnarray}
	\mathcal{G}_{ij}(\omega,\vec{x},\vec{x}') &=& 
	\int\frac{{\rm d}^3 k}{(2\pi)^3} \frac{{\rm d}^3 h}{(2\pi)^3}
	\mathcal{G}_{ij}(\omega,\vec{k},\vec{h})
	\,{\rm e}^{i(\vec{k}\cdot\vec{x}+\vec{h}\cdot\vec{x}')}~,
\label{eq:Green-Fourier}
\\
\mathcal{G}^*_{ji}(\omega,\vec{x}',\vec{x}) &=&
\int \frac{{\rm d}^3 k}{(2\pi)^3}\frac{{\rm d}^3 h}{(2\pi)^3}
	\mathcal{G}^*_{ji}(\omega,-\vec{h},-\vec{k})
	\,{\rm e}^{i(\vec{k}\cdot\vec{x}+\vec{h}\cdot\vec{x}')}~.
\label{eq:Green-star-Fourier}
\end{eqnarray}
In the frame where the field is in equilibrium and in the Dzyaloshinskii
gauge, we construct the covariant
Green tensor $\mathcal{G}_{\mu\nu}( k, h )$ from
\begin{equation}
	\mathcal{G}_{ij}( k, h ) =
	2\pi\delta( \omega + \omega' )
	\mathcal{G}_{ij}( \omega, \vec{k}, \vec{h} )
	~, \qquad
	\mathcal{G}_{0 \nu} = \mathcal{G}_{\mu 0} = 0
\label{eq:upgrade-Green}
\end{equation}
One can check straightforwardly that the field
correlations~(\ref{eq:DfdtEij}--\ref{eq:Dfdthij}) are equivalent to the 
following spectrum of the Faraday tensor 
\begin{eqnarray}
	\langle {F}_{\mu\nu}^{\rm fl}(k) , {F}_{\kappa\lambda}^{\rm fl}(h) \rangle_{F}
	= 
	\begin{aligned}[t]
	\frac{ {\rm i} \hbar }{ 2 }
	&
	\coth\frac{ \beta_F \cdot k }{ 2 } \times
\\
&
	\left[
	k_{[\mu} \mathcal{G}_{\nu][\kappa}( k, h ) h_{\lambda]}
	- h_{[\lambda} \mathcal{G}^{*}_{\kappa][\nu}( - h, - k) k_{\mu]}
	\right]
	~,
	\end{aligned}
\label{eq:DfdtFmunu}
\end{eqnarray}
where the brackets denote again antisymmetrized pairs of indices
[see Eq.(\ref{eq:anti-symmetrize})].
As in Eq.(\ref{eq:FD-dipole-cov}), we have expressed the inverse
temperature in terms of the 4-vector
$\beta_F^\mu = (\hbar / k_B T_F) u^\mu_F$ where $u^\mu_F$ is
the velocity of the field's equilibrium frame. 

With Eq.(\ref{eq:DfdtFmunu}), the fluctuation-dissipation
theorem for the fields
is now in manifestly covariant form. We emphasize that the only input
parameters required are the photon propagator and the 4-velocity of
the field's equilibrium frame. A general observer
notices the Doppler shift via the contraction $u^\mu_F k_\mu$
and a net energy flow (Poynting vector) parallel to $u^\mu_F$.
A general proof
that this expression is also gauge-invariant will be given elsewhere.
We show below that with a Green tensor $\mathcal{G}_{\mu\nu}$
in the generalized Coulomb gauge~\cite{Eberlein2006a}, we recover
the spectrum of the radiation force in the presence of a macroscopic
medium discussed in Refs.\cite{kyasov2002relativistic, dedkov2003relativistic}. 
We have also checked that in free space, the Green tensor
in the Feynman gauge [Eq.(\ref{eq:Feynman-Green-free-space})] 
reproduces from the fluctuation-dissipation theorem~(\ref{eq:DfdtFmunu})
the well-known expressions for the electric and magnetic field spectra.

As a side remark, we note that the fluctuation-dissipation 
theorems~(\ref{eq:FD-dipole-cov}, \ref{eq:DfdtFmunu}) display tensor
structures that are manifestly antisymmetric in the double indices
carried by the polarization and electromagnetic fields. The response
functions written in Eqs.(\ref{eq:4-rank-alpha}, \ref{eq:link-Green-tensors}) 
are only apparently of lower symmetry. They do preserve the parity of
the index pairs, however, so that an antisymmetric field induces an 
antisymmetric polarization, for example. But since the response functions are 
read off from the linear relations between antisymmetric quantities
[see, e.g., Eq.(\ref{eq:def-Green-4-tensor})],
their symmetric part actually remains undetermined. In other words,
we can also use in the fourth rank 
polarizability~(\ref{eq:4-rank-alpha}) a tensor with the structure
\begin{equation}
	\frac14 \left\{
	u^{[\mu} g^{\nu][\kappa} u^{\lambda]} + 
	u^{[\nu} g^{\mu][\lambda} u^{\kappa]}
	\right\} = 
	\frac{ \Gamma^{[\mu\nu][\kappa\lambda]} }{ 2 }
\end{equation}
so that the structural simplicity in the fluctuation-dissipation relation is preserved
even in the covariant formulation. (For a simplification of the field spectra,
see the planar geometry discussed below.)

\section{Calculation of the force}
\label{s:general-formula}

The two terms in the radiation force density~(\ref{eq:Dforcesplitting}) yield
a force acting on the atom,
combining the response functions with the fluctuation-dissipation theorems 
for field and polarization,
\begin{eqnarray}
	F &=& \int\!{\rm d}^3x \left\{ f^{(1)}( x ) + f^{(2)}( x ) \right\}
\label{eq:integrate-force}
\\
	f_\mu^{( 1 )}( x ) &=&
	\begin{aligned}[t]
      - i\int\frac{{\rm d}^4 k}{(2\pi)^4}\frac{{\rm d}^4 w}{(2\pi)^4}\frac{{\rm d}^4 h}{(2\pi)^4}
\,w_\eta&\alpha^{\eta\nu\kappa\lambda}(w,h)\times 
\\ 
&\left\langle {F}_{\mu\nu}^{\rm fl}(k){F}_{\kappa\lambda}^{\rm fl}(h)\right\rangle 
{\rm e}^{-i(k + w)\cdot x}
	\label{eq:Dforcesplittingfinalforceeins}
\end{aligned}
\\
	f_\mu^{( 2 )}( x ) &=&
	\begin{aligned}[t]
- i\int\frac{{\rm d}^4 k}{(2\pi)^4}\frac{{\rm d}^4 w}{(2\pi)^4}\frac{{\rm d}^4 h}{(2\pi)^4} 
\,w_\eta&\mathcal{G}_{\mu\nu\kappa\lambda}(k,-h)\times 
\\ 
&\left\langle{M}_{\rm fl}^{\kappa\lambda}(h){M}_{\rm fl}^{\eta\nu}(w)\right\rangle  
{\rm e}^{-i(k + w)\cdot x}~.
	\label{eq:Dforcesplittingfinalforcezwei}
\end{aligned}
\end{eqnarray}
The two contributions are worked out separately in Secs.~\ref{s:field-fluctuations}
and~\ref{s:dipole-fluctuations}. We specialize first the geometry of the field to
a stationary situation.

We focus on a planar geometry for the equilibrium field,
with the atom moving along a translation-invariant direction in the
$xy$-plane. We choose 
a frame where the field is in equilibrium, and denote spatial projections 
onto the $xy$-plane by the index $\Vert$. In this situation, 
translational
invariance entails that the Green tensor has the property 
\begin{eqnarray}
    \mathcal{G}_{\mu\nu}( k, h ) &=& 
    (2\pi)^3 \delta( k_0 + h_0 ) \delta( \vec{k}_\Vert + \vec{h}_\Vert )
    \mathcal{G}_{\mu\nu}( k, h_z )
\nonumber\\
    \mathcal{G}^{*}_{\nu\mu}( - h, - k ) &=& 
    (2\pi)^3 \delta( h_0 + k_0 ) \delta( \vec{h}_\Vert + \vec{k}_\Vert )
    \mathcal{G}^{*}_{\mu\nu}( - h, - k_z )
\label{eq:symmetry-Green}
\end{eqnarray} 
Note that this property holds in any inertial frame moving parallel to the
$xy$-plane. This leaves only the integration over $h_z$ in Eqs.(%
\ref{eq:Dforcesplittingfinalforceeins}, \ref{eq:Dforcesplittingfinalforcezwei}).
In addition, the spatial integration over $\vec{x}$ 
in Eq.(\ref{eq:integrate-force}) yields a 
$\delta( \vec{k} + \vec{w} )$.

\subsection{Field fluctuations}
\label{s:field-fluctuations}

In the equilibrium frame of the field, we have 
$\beta_F \cdot k = \beta k_0$. Inserting the correlation spectrum of 
$F_{\mu\nu}$ in Eq.(\ref{eq:Dforcesplittingfinalforceeins}) and performing
the simplifications mentioned above, the contribution of field
fluctuations to the force can be written in the form
($w^\mu = (w_0, -\vec{k})$, $h^\mu = (- k_0, -\vec{k}_\Vert, h_z)$)
\begin{eqnarray}
    F_\mu^{(1)} &=& \frac{ \hbar }{ 2 }
\int\!\frac{ {\rm d}^4k }{ (2\pi)^4 }
    \frac{ {\rm d}w_0 }{ 2\pi }
    \frac{ {\rm d}h_z }{ 2\pi }
    \coth \!\big( \frac{ \beta k_0 }{ 2 } \big)
    w_\eta \alpha^{\eta\nu\kappa\lambda}( w, h )
\times
\nonumber\\
    &&
    \qquad
    \left[
    k_{[\mu} \mathcal{G}_{\nu] [\kappa}( k, h_z ) h_{\lambda]}
    -
    h_{[\lambda} \mathcal{G}^*_{\kappa][\nu}( - h, - k_z ) k_{\mu]}
    \right]
\,{\rm e}^{ - {\rm i} (w_0 + k_0) t }
\label{eq:force-eins-1}
\end{eqnarray}
The polarizability $\alpha^{\eta\nu\kappa\lambda}( w, h )$ 
[Eq.(\ref{eq:4-rank-alpha})] contains a $\delta$-function
whose argument becomes for an atom with velocity 
$u^\mu = \gamma ( 1, \vec{v} )$ in the $xy$-plane: 
\begin{equation}
	u\cdot(w - h) = \gamma (w_0 + k_0) +
\gamma \vec{v} \cdot (\vec{k}_\Vert - \vec{k}_\Vert) = 
\gamma (w_0 + k_0)
~. 
	\label{eq:fix-frequency}
\end{equation}
This allows to perform the $w_0$ integral
and completes the identification $w = - k$. 
We are therefore justified to work with the following expression 
for the polarizability tensor
\begin{equation}
      \alpha^{\eta\nu\kappa\lambda}( w, h ) =
      2\pi 
      \frac{ \delta( w_0 + k_0 ) }{ \gamma }
      \alpha( - u \cdot k )
      \left\{ u^{\eta} g^{\nu\kappa} u^{\lambda} 
      + u^{\nu} g^{\eta\lambda} u^{\kappa} \right\}
	\, {\rm e}^{ {\rm i} ( k_z + h_z ) z_A }
\label{eq:alpha-4-simplified}
\end{equation}
As expected, the force resulting from Eq.(\ref{eq:force-eins-1})
is constant in time and depends only on the 
atom-surface distance $z_A$.

The contraction of the two terms in brackets 
in Eq.(\ref{eq:force-eins-1}) with the fourth rank tensor in the 
polarizability~(\ref{eq:alpha-4-simplified}) gives
\begin{eqnarray}
&&
- k_\eta \left\{ 
      u^{\eta} g^{\nu\kappa} u^{\lambda} + u^{\nu} g^{\eta\lambda} u^{\kappa}
\right\}
k_{[\mu} \mathcal{G}_{\nu] [\kappa}( k, h ) h_{\lambda]}
= k_\mu \phi( k, h )
\\
&& 
- k_\eta \left\{ 
      u^{\eta} g^{\nu\kappa} u^{\lambda} + u^{\nu} g^{\eta\lambda} u^{\kappa}
\right\}
      h_{[\lambda} \mathcal{G}^*_{\kappa][\nu}( - h, - k ) k_{\mu]}
= k_\mu \bar\phi( -h, -k )
~,
\label{eq:def-phi-bar}
\end{eqnarray}
where we defined the scalar function
\begin{eqnarray}
&& \phi( k, h )
= 
      - (u \cdot k) (u \cdot h) {\mathcal{G}^\kappa}_{\kappa}(k, h)  
      - (k \cdot h) u^\nu u^\kappa \mathcal{G}_{\nu\kappa}(k, h)  
\nonumber
\\
&&
\hphantom{\phi( k, h ) = }
      + (u \cdot h) u^\nu k^\kappa \mathcal{G}_{\nu\kappa}(k, h)
      + (u \cdot k) h^\nu u^\kappa \mathcal{G}_{\nu\kappa}(k, h) 
\label{eq:Dfirstfirstcontraction}
\end{eqnarray}
The function $\bar\phi( -h, -k )$ in Eq.(\ref{eq:def-phi-bar})
is obtained by the replacement
$\mathcal{G}_{\nu\delta}(k, h) \mapsto  \mathcal{G}^*_{\delta\nu}( -h, - k )$
in the definition~(\ref{eq:Dfirstfirstcontraction}) of $\phi( k, h )$. We note
that this formula is valid for arbitrary $k$ and $h$.

We end up with the following integral for this piece of the radiation force
($k^\mu = (\omega, \vec{k}_\Vert, k_z)$
and $h^\mu = (-\omega, -\vec{k}_\Vert, h_z)$)
\begin{equation}
     F_\mu^{(1)} =
\frac{ \hbar }{ 2 }
\int\!\frac{ {\rm d}^4k }{ (2\pi)^4 }
      \frac{ {\rm d}h_z }{ 2\pi }
      \coth \!\big( \frac{ \beta_F \omega }{ 2 } \big)
      \frac{ k_\mu }{ \gamma }
      \alpha( - u \cdot k )
    \left[ \phi( k, h ) 
      - \bar\phi( -h, -k)
      \right]
\, {\rm e}^{ {\rm i} ( k_z + h_z ) z_A }
\label{eq:force-eins-2}
\end{equation}
This formula and Eq.(\ref{eq:force-zwei-1}) below, giving the two pieces 
in $F_\mu$, are our main result.

\subsection{Dipole fluctuations}
\label{s:dipole-fluctuations}

The contribution of dipole fluctuations [Eq.(\ref{eq:Dforcesplittingfinalforcezwei})]
is worked out in a similar way. 
We find again that only $w = - k$ is relevant. 
The contraction of the fourth rank tensors
involves the remarkable identity
\begin{equation}
	- k_\eta \left\{
	k_{\mu} \mathcal{G}_{\nu\kappa}( k, - h ) h_{\lambda} + 
	k_{\nu} \mathcal{G}_{\mu\lambda}( k, - h ) h_{\kappa}
	\right\}
	u^{[\kappa} g^{\lambda] [\eta} u^{\nu]}
	=
	k_\mu \Phi( k, - h )		
	\label{eq:amazing-identity}
\end{equation}
where the function $\Phi( k, - h )$ has the same definition as
$\phi( k, h )$ in Eq.(\ref{eq:Dfirstfirstcontraction}), except for 
the replacement $\mathcal{G}_{\nu\delta}(k, h) \mapsto
\mathcal{G}_{\nu\delta}(k, - h)$.
The force due to dipole fluctuations finally takes the form
(here, $h = (k_0, \vec{k}_\Vert, h_z)$)
\begin{equation}
	F_\mu^{( 2 )} =
- {\rm i}\hbar 
\int\frac{{\rm d}^4 k}{(2\pi)^4} 
\frac{{\rm d} h_z}{ 2\pi } 
\,\coth \!\big( \frac{ \beta_A \cdot k }{ 2 } \big)
\frac{ k_\mu }{ \gamma } 
\,{\rm Im}\,\alpha( u \cdot k )
\Phi( k, -h )
\,{\rm e}^{i(k_z - h_z)z_A}
~,
\label{eq:force-zwei-1}
\end{equation}
which has a structure quite similar to $F_\mu^{(1)}$
[Eq.(\ref{eq:force-eins-2})].

\subsection{Blackbody friction}
\label{s:bb-friction}

As an illustration and check of this approach, we consider two simple
situations: motion in blackbody radiation (this Section) and 
above a planar dielectric (Sec.\ref{s:surface-friction}). In the first case,
we use the free-space photon Green's 
function~(\ref{eq:Feynman-Green-free-space}) 
in the Feynman gauge, in the second case, the Green function obtained
in Ref.\cite{Eberlein2006a} in a generalized Coulomb gauge. Recall
that friction forces
in the blackbody radiation field have been studied in the early days of
quantum theory by 
Einstein \& Hopf \cite{Einstein1910} and Einstein \cite{Einstein1917}.

The full translational symmetry 
of the photon propagator~(\ref{eq:Feynman-Green-free-space}) 
entails the additional $\delta$-function
$\delta( k_z + h_z )$ in $\mathcal{G}_{\mu\nu}( k, h )$
[Eq.(\ref{eq:symmetry-Green})]. Using the resulting values for the
4-vector $h$, the scalar functions in the two pieces
$F^{(1)}$ and $F^{(2)}$ are worked out to be
\begin{eqnarray}
	\phi( k, h ), \, \bar\phi( -h, -k ) &\mapsto & 
	- \frac{ 2 (u \cdot k)^2 + k^2 }{ k^2 \pm {\rm i} 0 
	\mathop{\mathrm{sgn}}( \omega ) }
	\label{eq:scalar-fs-eins}
\\
	\Phi( k, -h ) & \mapsto &
	\frac{ 2 (u \cdot k)^2 + k^2 }{ k^2 + {\rm i} 0 
	\mathop{\mathrm{sgn}}( \omega ) }
\end{eqnarray}
The combination $\phi( k, h ) - \bar\phi( -h, -k )$ becomes proportional
to a $\delta$-function localized on the light cone $k^2 = 0$, thus removing 
the term $k^2$ in the numerator. The integration 
over the direction of $\vec{k}$ remains:
we observe that only the components $F_0$ and $F_x$ are nonzero
(due to parity) in a frame where $\vec{v}$ is along the $x$-axis.
Under the reflection $( \omega, k_x ) \mapsto ( -\omega, -k_x )$ 
that flips the sign of $u \cdot k$, only the imaginary part
${\rm Im}\,\alpha( - u \cdot k ) = - {\rm Im}\,\alpha( u \cdot k )$
gives an even integrand in Eq.(\ref{eq:force-eins-2}). 
Pulling these facts together, we find
(solid angle ${\rm d}\Omega$ for unit vector $\vec{k} / |\omega|$)
\begin{equation}
	F_\mu^{(1)} = 
	\frac{ \hbar }{ 2\pi \gamma }
	\int\limits_{-\infty}^\infty\!\frac{ {\rm d}\omega }{ 2\pi }
	\int\!\frac{ {\rm d}\Omega }{ 4\pi }
	\coth \!\big( \frac{ \beta_F \omega }{ 2 } \big)
	\omega k_\mu ( u \cdot k )^2 
	\,{\rm Im}\,\alpha( u \cdot k )	
	\label{eq:force-eins-3}
\end{equation}
For the second contribution
\begin{equation}
	F_\mu^{(2)} = 
- {\rm i}\hbar \int\frac{{\rm d}^4 k}{(2\pi)^4}
\coth \!\big( \frac{ \beta_A \cdot k }{ 2  } \big)
\frac{ k_\mu }{ \gamma } 
\,{\rm Im}\,\alpha( u \cdot k )
\frac{ 2 (u \cdot k)^2 + k^2 }{ k^2 + {\rm i} 0 
	\mathop{\mathrm{sgn}}( \omega ) }
~,
	\label{eq:force-zwei-2}
\end{equation}
we make the integrand even under the reflection mentioned above
Eq.(\ref{eq:force-eins-3})
and get again an $- 2 \pi {\rm i} 
\mathop{\mathrm{sgn}}( \omega ) \delta( k^2 )$ from the last term.
The integral then reduces to
\begin{equation}
	F_\mu^{(2)} = 
- \frac{ \hbar }{ 2 \pi \gamma } 
\int\limits_{-\infty}^\infty\!\frac{{\rm d} \omega}{ 2\pi}
\int\!\frac{{\rm d}\Omega}{4\pi}
\coth \!\big( \frac{ \beta_A \cdot k }{ 2  } \big)
\omega k_\mu 
(u \cdot k)^2
\,{\rm Im}\,\alpha( u \cdot k )
~,
	\label{eq:force-zwei-3}
\end{equation}
which has nearly the same structure as $F^{(1)}$ [Eq.(\ref{eq:force-eins-3})]
except that 
$\coth( \beta_A \cdot k / 2 )$ involves the atomic temperature
and the Doppler-shifted frequency $\omega'_A = u\cdot k$ in the 
atom's rest frame.

The total force thus features the difference
\begin{equation}
	\coth\frac{ \beta_A \cdot k }{ 2  } - 
	\coth\frac{ \beta_F \omega }{ 2  } = 
	2 N( \omega'_A, T_A ) - 2 N( \omega, T_F )
	\label{eq:difference-BE-occupations}
\end{equation}
where $N( \omega, T )$ is the Bose-Einstein distribution. (The co-moving
frequency $\omega'_A$ has necessary the same sign as $\omega$
for field modes on the light cone.) 
This makes the frequency integration converge exponentially fast
at $|\omega| \to \infty$. And it is easy to check that the sum of 
Eqs.(\ref{eq:force-eins-3}, \ref{eq:force-zwei-3}) is equal to the
radiative friction force calculated in 
Refs.\cite{Mkrtchian2003, kyasov2002relativistic} for the case of a small 
polarizable particle (for a review and discussion, see
Ref.\cite{Dedkov2010a, Intravaia2011}).

\subsection{Friction above a dielectric surface}
\label{s:surface-friction}

We finally address the situation that a neutral particle is
moving parallel to a dielectric surface where the radiation force acts
as friction. This issue has recently received
some regain of interest, in particular after the claim in 
Ref.\cite{philbin2009no} that for two macroscopic bodies in relative
motion, the frictional stress should vanish for $T \to 0$. 
(See Refs.\cite{Volokitin2009comment,%
Pendry2010comment,Dedkov2010comment} for further discussion.)

\subsubsection*{Reflected photon Green function.}

The starting point is the photon Green function $\mathcal{G}_{\mu\nu}( x, 
x' )$ for which we take the expression derived in Ref.\cite{Eberlein2006a}
(both $z, z' > 0$, outside the surface)
\begin{eqnarray}
	\mathcal{G}_{\mu\nu}( x, x' ) &=& 
	- \int\!\frac{ {\rm d}k_0 }{ 2\pi }
	\frac{ {\rm d}^2 k_\Vert }{ (2\pi)^2 }
	{\rm e}^{ - {\rm i} 
	[
	k_0 ( t - t' ) -
	\vec{k}_\Vert \cdot ( \vec{x} - \vec{x}' )
	]}
	\times
\nonumber
\\
&&	
	\Big\{
	\int\!\frac{ {\rm d}k_z }{ 2\pi }
	\frac{ g_{\mu\nu}\,
	{\rm e}^{ {\rm i} k_z (z - z') }
	}{ k^2 + {\rm i} 0 \mathop{\mathrm{sgn}}{\omega} }
+
	\int_{\cal C}\!\frac{ {\rm d}k_z }{ 2\pi }
	\sum_\sigma 
	\frac{ r_\sigma P^{(\sigma)}_{\mu\nu}
	\, {\rm e}^{ {\rm i} k_z (z + z') }
	}{ k^2 + {\rm i} 0 \mathop{\mathrm{sgn}}{\omega} }
	\Big\}
\label{eq:Feynman-Green-Eberlein}
\end{eqnarray}
where $\sigma$ is a polarization index and the quantities
$r_\sigma$,  $P^{(\sigma)}_{\mu\nu}$ are detailed in 
Eq.(\ref{eq:define-polarizations}) below. This is written in the rest frame
of the dielectric medium. The first term in curly
brackets is the same as in free space 
[Eq.(\ref{eq:Feynman-Green-free-space})], and we can focus here
on the second term. It is built from waves that are reflected from the
planar surface, as illustrated by the sign flip in front of the
second coordinate $z'$. The integral 
is over a contour $\mathcal{C}$ in the complex $k_z$ plane, including the real 
axis and 
two segments running on opposite sides of the imaginary axis.
In our notation where $z, z' > 0$, these segments are between $k_z = 0$ and 
$k_z = {\rm i} |\vec{k}_\Vert| \sqrt{ 1 - 1/n^2 }$ where $n > 1$ is the
refractive index of the dielectric medium below the surface~\cite{Eberlein2006a}.
The reflection matrices involve two transverse polarizations
$\sigma = s, p$ which are gauge-independent, and two gauge-dependent
ones, scalar $\sigma = l$ and longitudinal $\sigma = k$:
\begin{equation}
\begin{aligned}
	r_s & = \frac{k_z - k'_z}{k_z + k'_z}~,
	&
	P^{(s)}_{ij} & =
	\begin{pmatrix}
		-k_y^2 & k_x k_y & 0 \\
		k_x k_y & -k_x^2 & 0 \\
		0 & 0& 0
	\end{pmatrix}
	\frac{1}{\vec{k}_\Vert^2}~,
\\
	r_p & = \frac{n^2 k_z - k'_z}{n^2 k_z + k'_z}~,
	&
	P^{(p)}_{ij} & =
	\begin{pmatrix}
		k_z^2 k_x^2 & k_z^2 k_x k_y  & k_z k_x \vec{k}_\Vert^2\\
		k_z^2 k_x k_y & k_z^2 k_y^2 & k_z k_y \vec{k}_\Vert^2 \\
		-k_z k_x \vec{k}_\Vert^2  &  -k_z k_y \vec{k}_\Vert^2 
				& -\vec{k}_\Vert^4
	\end{pmatrix}
	\frac{1}{\vec{k}_\Vert^2\vec{k}^2}~,
\\
	r_l & = \frac{k_z - n^2 k'_z}{k_z + n^2 k'_z}~,
	&
	P^{(l)}_{00} & =  1	
\\
	r_k & = \frac{k_z - n^2 k'_z}{k_z + n^2 k'_z}~,
	&
	P^{(k)}_{ij} & =
	\begin{pmatrix}
		-k_x^2 & -k_x k_y & k_x k_z \\
		-k_x k_y & -k_y^2 & k_y k_z \\
		-k_x k_z & -k_y k_z & k_z^2
	\end{pmatrix}
	\frac{1}{\vec{k^2}}~.
\end{aligned}
	\label{eq:define-polarizations}
\end{equation}
All other components vanish. (The gauge chosen 
in Ref.\cite{Eberlein2006a} is a generalized Coulomb one.)
Note that the scalar and longitudinal polarizations
have the same reflection amplitude $r_l = r_k$.
Finally, the reflection amplitudes involve the medium wavevector 
$k_z'$ given by
\begin{equation}
	k_z' = \sqrt{ n^2 k_z^2 + (n^2 - 1) \vec{k}_\Vert^2 }
	\label{eq:def-medium-wavevector}
\end{equation}
where the square root must be evaluated with a branch cut joining
the points $k_z = - {\rm i} |\vec{k}_\Vert| \sqrt{ 1 - 1/n^2 }$
and $k_z = + {\rm i} |\vec{k}_\Vert| \sqrt{ 1 - 1/n^2 }$.
The integration contour ${\cal C}$ avoids this branch cut from above.
% identify $h_z$ integration, perform $k_z$ integral with contour tricks
In the following, we denote by $\mathcal{R}$ the reflected part
of the photon Green function and keep only this contribution.
From the expression~(\ref{eq:Feynman-Green-Eberlein}) of 
$\mathcal{R}( x, x' )$, we read
off that the double Fourier representation used before in 
Eq.(\ref{eq:Green-Fourier}) has the property
\begin{equation}
	\mathcal{R}_{\mu\nu}( k , h ) = 
    (2\pi)^4 \delta( k_0 + h_0 ) \delta( \vec{k}_\Vert + \vec{h}_\Vert )
    \delta( k_z - h_z ) \mathcal{R}_{\mu\nu}( k )
\label{eq:symmetry-Green-reflected}
\end{equation}
so that $h = - k_r = - (k_0, \vec{k}_\Vert, - k_z)$ is fixed to the reflected 
wave vector.

We have checked that this formulation (with the contour integral over 
$k_z$ and the relation~(\ref{eq:symmetry-Green-reflected}) fixing $h_z$)
carries through the previous calculation in place of the ordinary 
Fourier integrals over $k_z$ and $h_z$. A similar expansion also holds for
the conjugate tensor ${\cal R}^*$. Indeed, using the fact that the contour
is mapped according to $\mathcal{C}^* = - \mathcal{C}^{-1}$, and the
properties of the
scattering amplitudes under complex conjugation compiled in 
Ref.\cite{Eberlein2006a}, one can convince oneself that the conjugate
tensor $[\mathcal{R}_{\nu\mu}( x', x )]^*$ can be written exactly
as the second term in Eq.(\ref{eq:Feynman-Green-Eberlein}), except
that the retarded denominator must be replaced by the advanced one,
$k^2 - {\rm i} 0 
\mathop{\mathrm{sgn}}( \omega ) $.

\subsubsection*{Reduction to the light cone.}

We perform the integration over $k_z$ by closing
the contour ${\cal C}$ with a half-circle at infinity in the upper half-plane
(observe ${\rm e}^{ {\rm i} k_z (z + z') }$ in Eq.(\ref{eq:Feynman-Green-Eberlein})). There are two poles, one at $k_z = {\rm i} | \vec{k}_\Vert|$
from the normalization factor $\vec{k}^2$ in the projectors,
and another one at 
\begin{equation}
	k_z = {\rm i} \sqrt{ \vec{k}_\Vert^2 - (\omega + {\rm i} 0)^2 }
		\equiv {\rm i} \kappa
	\label{eq:kz-on-light-cone}
\end{equation}
from the photon propagator (on the light cone). It is easy to check that 
the residues
at the former pole compensate between the $p$- and longitudinal
polarizations, the reflection coefficients taking the values
$r_p = (n^2 - 1)/(n^2 + 1) = - r_k$. For more technical details,
see Ref.\cite{Pieplow2012qfext}.

Another cancellation happens between 
the longitudinal and scalar polarizations on the light cone when the scalar function
$\phi( k, h )$ [Eq.(\ref{eq:Dfirstfirstcontraction})] is evaluated.
Let us write $\phi_\sigma( k )$ ($\sigma = l, k$) for the 
corresponding expressions when the projectors $P^{(\sigma)}_{\mu\nu}$
are replaced for $\mathcal{R}_{\mu\nu}( k, h )$ 
(we use $h = - k_r$):
\begin{equation}
	\phi_\sigma( k ) = 
      (u \cdot k) (u \cdot k_r) {P^{(\sigma)\lambda}}_{\lambda}
      + (k \cdot k_r) u^\nu u^\lambda P^{(\sigma)}_{\nu\lambda}
      - (u \cdot k_r) u^\nu k^\lambda P^{(\sigma)}_{\nu\lambda}
      - (u \cdot k) k_r^\nu u^\lambda P^{(\sigma)}_{\nu\lambda}
\label{eq:def-polarization-weight-phi-sigma}
\end{equation}
The scalar polarization picks the time-like components, while the longitudinal
polarization projects onto $\vec{k}$ and $\vec{k}_r$. Indeed, 
the latter projector is re-written (on the light cone) as
\begin{equation}
	P^{(k)}_{ij} = - \frac{ k_i k_{rj} }{ \vec{k}^2 } = 
		- \frac{ k_i k_{rj} }{ \omega^2 }
	\label{eq:longitudinal-projector}
\end{equation}
where $\vec{k}_r = (\vec{k}_\Vert, -k_z)$ is the reflected wave vector.
Straightforward 
algebra shows that
\begin{equation}
k^2 = 0 \quad \Rightarrow \quad
	r_l \phi_l( k ) + r_k \phi_k( k ) = 0
	\label{eq:cancellation}
\end{equation}
This is again an indication that our covariant expression for the radiation
force is gauge invariant.

The conjugate Green tensor ${\cal R}^*$ is handled in a similar way, 
taking care of the positions of the poles in $k_z$. The sum of the two terms 
in $F^{(1)}$ [Eq.(\ref{eq:force-eins-2})] results in:
\begin{eqnarray}
	&& \int_{\cal C}\!\frac{ {\rm d}k_z }{ 2\pi }\!\frac{ {\rm d}h_z }{ 2\pi }
	k_\mu \left[
	\phi( k, h ) - \bar \phi( -h, -k ) 
	\right]
	{\rm e}^{ {\rm i} (k_z + h_z) z_A }
\nonumber
\\
	&& \qquad =
\frac{ 1 }{ 2 }
\sum_{\sigma \,= \,s,p}
\phi_\sigma( k )
\left[
k_\mu
\frac{ r_\sigma \, {\rm e}^{ - 2 \kappa z_A } }{
\kappa }
-
\bar k_\mu
\frac{ r_\sigma^*  \, {\rm e}^{ - 2 \kappa^* z_A } }{
\kappa^*
 }
\right]
	\label{eq:simplify-force-eins}
\end{eqnarray}
where the light-like vectors $k^\mu$ and $\bar k^\mu$ have $z$-components
given by ${\rm i}\kappa$ [Eq.(\ref{eq:kz-on-light-cone})] and
${\rm i}\kappa^*$, respectively. The polarizations come with real-valued
weight functions
\begin{eqnarray}
	\phi_s( k ) &=& % ( u \cdot k )^2 
	\gamma^2 ( \omega - \vec{v} \cdot \vec{k}_\Vert )^2
	+ 2 \gamma^2 (\vec{v} \times \vec{k}_\Vert)^2 
	\left( 1 - \frac{ \omega^2 }{ \vec{k}_\Vert^2 } \right)
\\
	\phi_p( k ) &=& % ( u \cdot k )^2 
	\gamma^2 ( \omega - \vec{v} \cdot \vec{k}_\Vert )^2
	+ 2 \gamma^2 ( \vec{k}_\Vert^2 - (\vec{v} \cdot \vec{k}_\Vert)^2 ) 
	\left( 1 - \frac{ \omega^2 }{ \vec{k}_\Vert^2 } \right)
\end{eqnarray}

\subsubsection*{Field fluctuations.}

For comparison 
with Ref.\cite{dedkov2003relativistic}, we work out the friction force $F_x$
parallel to $\vec{v}$. The terms in brackets in 
Eq.(\ref{eq:simplify-force-eins}) are then complex conjugates one of the other.
From Eq.(\ref{eq:kz-on-light-cone}) for $\kappa$ and the properties of the
reflection coefficients in Ref.\cite{Eberlein2006a}, we observe that this function 
is odd under a sign flip of both $\omega$ and $k_x$. Keeping only even terms
in the integrand, we end up from Eqs.(\ref{eq:force-eins-2}) 
and~(\ref{eq:simplify-force-eins}) with the manifestly real expression for 
that part of the force that depends on the field temperature
\begin{equation}
	F_x^{(1)} =
	\frac{ \hbar }{ 2 \gamma }
\int\!\frac{ {\rm d}\omega }{ 2\pi }
\frac{ {\rm d}^2k_\Vert }{ (2\pi)^2 }
      \coth \!\big( \frac{ \beta_F \omega }{ 2 } \big)
	k_x
	\, {\rm Im}\,
      \alpha( u \cdot k ) \!
      \sum_{\sigma \,= \,s,p}
      \phi_\sigma( k )
      \, {\rm Im}\left(
\frac{ r_\sigma  \, {\rm e}^{ - 2 \kappa z_A } }{
\kappa }
\right)
\end{equation}

\subsubsection*{Dipole fluctuations.}

The integral over dipole fluctuations~(\ref{eq:force-zwei-1}) is similar.
The Green function ${\cal R}_{\mu\nu}( k, -h )$ [Eq.(\ref{eq:symmetry-Green-reflected})] 
involves $\delta( k_r - h )$ and fixes $h$. Performing the $k_z$-integration,
we get
\begin{equation}
h = k_r: \quad
\int_{\cal C}\!\frac{ {\rm d}k_z }{ 2\pi }\!\frac{ {\rm d}h_z }{ 2\pi }
	k_\mu 
	\Phi( k, - h )
	{\rm e}^{ {\rm i} (k_z - h_z) z_A }
= \sum_{\sigma}
	(- \phi_\sigma( k )) \frac{ r_\sigma \,
	{\rm e}^{ - 2 \kappa z_A } }{ 2 \kappa }
\end{equation}
where the weight functions defined in Eq.(\ref{eq:def-polarization-weight-phi-sigma})
appear again. 
Putting this into Eq.(\ref{eq:force-zwei-1}) and picking the
even part of the integrand, we get
\begin{equation}
	F_\mu^{(2)} =
	- \frac{ \hbar }{ 2 \gamma }  
\int\frac{ {\rm d}\omega }{ 2\pi } 
\frac{ {\rm d}^2 k_\Vert }{ (2\pi)^2 } 
\,\coth \!\big( \frac{ \beta_A \cdot k }{ 2 } \big)
k_\mu
\,{\rm Im}\,\alpha( u \cdot k )
\sum_{\sigma}
	\phi_\sigma( k ) 
	\,{\rm Im}\left(
	\frac{ r_\sigma \,{\rm e}^{ - 2  \kappa z_A }		}{ 
	\kappa }
	\right)
\end{equation}
The net force thus involves the same difference of thermal occupations
as in free space [Eq.(\ref{eq:difference-BE-occupations})]. 

\subsubsection*{Comparison to previous results}

\begin{table}[bt]
\begin{center}\small
\begin{tabular}{| l | l l l l l l l l l l @{}|}
\hline
& 
\makebox[1.5em][l]{\rotatebox{70}{particle temperature}} &
\makebox[1.5em][l]{\rotatebox{70}{field temperature}} &
\makebox[1.5em][l]{\rotatebox{70}{particle velocity}} &
\makebox[1.5em][l]{\rotatebox{70}{(electric) polarizability }} &
\makebox[1.5em][l]{\rotatebox{70}{co-moving frequency}} &
\makebox[1.5em][l]{\rotatebox{70}{parallel wave vector}} &
\makebox[1.5em][l]{\rotatebox{70}{propagating waves}} &
\makebox[1.5em][l]{\rotatebox{70}{evanescent waves}} &
\makebox[1.5em][l]{\rotatebox{70}{reflection amplitude}} &
\makebox[1.5em][l]{\rotatebox{70}{polarization weight}}\hspace*{2.25em}
\\
\hline
\rule{0pt}{2.5ex}%
Ref.\cite{Dedkov2008} &
$T_1$ & $T_2$ & $\beta c, V$ & $4\pi\alpha_{\rm e}$
	& $\gamma \omega^-$ & ${\bf k}$ & $\tilde q_0$
	& ${\rm i}q_0$ & $\tilde\Delta_{\rm m, e}$, $\Delta_{\rm m,e}$
	& $\gamma^2 \chi^-_{\rm m,e}$
\\
this paper &
$T_A$ & $T_F$ & $v_x$ & $\alpha$ 
	& $u \cdot k$ & $\vec{k}_\Vert$ & $k_z \in \mathbbm{R}$
	& $k_z = {\rm i}\kappa$ & $r_{s, p}$
	& $\phi_{s,p}$
\\
\hline
\end{tabular}
\end{center}
\caption[]{Dictionary of symbols. In Ref.\cite{Dedkov2008}, 
Gauss units are used.}
\label{t:translate}
\end{table}

In Refs.\cite{kyasov2002relativistic,dedkov2003relativistic}, the same problem 
was treated in a not manifestly covariant way. We compare here to 
Eq.(13) in the review~\cite{Dedkov2008} which is the sum of the friction force in free space [Sec.\ref{s:bb-friction}] and above a magneto-dielectric surface.
The free-space piece is equal to the sum of
Eqs.(\ref{eq:force-eins-3}, \ref{eq:force-zwei-3}), as can be shown with
the translation table~\ref{t:translate}.
We have checked that also the surface friction is the same as our result,
in both propagating and evanescent sectors 
($k_\Vert < \omega/c$ and $k_\Vert > \omega/c$, respectively),
taking care of the symmetry of the integrand with respect to the
signs of $\omega$ and $k_x$.
In Ref.
\cite{Scheel2009moving}, the scenario for calculating the force is slightly
different because the parameters in the atomic polarizability
(frequency shift, linewidth) are modified themselves
due to the interaction with the surface. We assume here that $\alpha( \omega )$
is an independent input parameter. Alternatively, one may work
with a ``dressed polarizability'' that depends, in general, on the atom-surface
distance.  

In Ref.\cite{volokitin2008theory}, 
the near-surface change in the polarizability is emphasized as well, and 
contributions up to order ${\cal O}[ \alpha^2( \omega ) ]$
are calculated. The starting point for radiative friction 
is not Eq.(\ref{eq:4-force-and-current}) for the (average)
force. A friction coefficient to linear order in the velocity $v$ is calculated 
from a 
force autocorrelation function, essentially similar to the Kubo 
formula in Eq.(\ref{eq:Kubo-Green-tensor}). The terms to lowest order
in the polarizability $\alpha$ and for a common temperature coincide
with the result of Ref.\cite{dedkov2003relativistic} and therefore with
ours.

\subsubsection*{Normal force.}

For completeness, we also give here the two contributions to the
normal component of the radiation force. This provides the generalization
of the well-known Casimir-Polder interaction to the situation of a moving
particle at a different temperature than the surface. The calculations are
the same and lead to 
\begin{eqnarray}
	\hspace*{-15mm}
     F_z^{(1)} &=&
- \frac{ \hbar }{ 2 \gamma }
\int\!\frac{ {\rm d}\omega }{ 2\pi }
\int\!\frac{ {\rm d}^2k_\Vert }{ (2\pi)^2 }
      \coth \!\big( \frac{ \beta_F \omega }{ 2 } \big)
      \,{\rm Re}\, \alpha( u \cdot k )
\sum_{\sigma \,= \,s,p}
\phi_\sigma( k )
\,{\rm Im}\left[
r_\sigma \, {\rm e}^{ - 2 \kappa z_A } 
\right]
\label{eq:normal-force-eins}
\\
	\hspace*{-15mm}
     F_z^{(2)} &=&
- \frac{ \hbar }{ 2 \gamma }
\int\!\frac{ {\rm d}\omega }{ 2\pi }
\int\!\frac{ {\rm d}^2k_\Vert }{ (2\pi)^2 }
      \coth \!\big( \frac{ \beta_A \cdot k }{ 2 } \big)
      \,{\rm Im}\, \alpha( u \cdot k )
\sum_{\sigma \,= \,s,p}
\phi_\sigma( k )
\,{\rm Re}\left[
r_\sigma \, {\rm e}^{ - 2 \kappa z_A } 
\right]
\label{eq:normal-force-zwei}
\end{eqnarray}
Both contributions are manifestly real. We have checked that these
results coincide with Eq.(12) of Ref.\cite{Dedkov2008} for both propagating
and evanescent modes.

\section{Conclusions}
\label{s:conclusion}

The problem of radiative friction on neutral particles near macroscopic
bodies or between two such objects has been addressed by several
authors in recent years, using different approaches. We have constructed 
here a framework that embodies several of these results, and has the advantage
of being manifestly compatible with the requirements of special relativity. 
The formulation highlights the different geometric objects that are involved
in the electromagnetic coupling, the material polarization is for example
an antisymmetric rank-two tensor, conjugate to the Faraday tensor.
We believe that one advantage of the formulation is to expound clearly 
the concept of local thermodynamic equilibrium which is a prerequisite to
apply the fluctuation-dissipation theorem in the relativistic context. From
this viewpoint, the two-temperature situations that have been studied
quite intensively over the previous years, appear on the same footing 
as two objects in relative motion.

We found hints that the radiative force on the particle is a gauge-independent
quantity, by retrieving previous results from different choices for the
relativistic photon propagator (Green tensor). It is also interesting that 
the covariant formulation displays the force [Eqs.(\ref{eq:force-eins-2},
\ref{eq:force-zwei-1})] as being proportional to the 4-wavevector 
$k_\mu$. This may help to interpret the associated potential energy 
in a covariant way and to compare with other results.

\paragraph*{Acknowledgments.}
We thank H. R. Haakh and V. E. Mkrtchian for valuable comments.
This work was supported by \emph{Deutsche Forschungsgemeinschaft}
(grant He-2849/4).

\end{document}